\documentstyle[11pt,epsf]{article}

\newcommand{\beq}{\begin{equation}} \newcommand{\eeq}{\end{equation}}
\newcommand{\bea}{\begin{eqnarray}} \newcommand{\eea}{\end{eqnarray}}

  \newcommand
{\Romannumeral}[1]{\uppercase\expandafter{\romannumeral#1}}

\newcommand{\be}{\begin{enumerate}} \newcommand{\ee}{\end{enumerate}}
\newcommand{\bi}{\begin{itemize}} \newcommand{\ei}{\end{itemize}}
\newcommand{\ba}{\begin{array}} \newcommand{\ea}{\end{array}}
\newcommand{\bc}{\begin{center}} \newcommand{\ec}{\end{center}}
\newcommand{\bt}{\begin{tabular}} \newcommand{\et}{\end{tabular}}

\def\lsim{\mathrel{\rlap{\lower4pt\hbox{\hskip1pt$\sim$}}
    \raise1pt\hbox{$<$}}}           
\def\gsim{\mathrel{\rlap{\lower4pt\hbox{\hskip1pt$\sim$}}
    \raise1pt\hbox{$>$}}}           

\newcommand{\half}{\textstyle {1\over2} \displaystyle}    
\newcommand{\Dslash}{{\hbox{D}\kern-0.6em\raise0.15ex\hbox{/}}} 

\renewcommand{\et}{\eta}

\begin{document}

\setlength{\oddsidemargin}{0cm} \setlength{\baselineskip}{7mm}

\input epsf

\begin{normalsize}\begin{flushright}

CERN-PH-TH/2006-147 \\
July 2006 \\

\end{flushright}\end{normalsize}

\begin{center}
  
\vspace{5pt}

{\Large \bf Constraints on Gravitational Scaling Dimensions}

{\Large \bf from Non-Local Effective Field Equations}

\vspace{40pt}

{\sl H. W. Hamber}
$^{}$\footnote{On leave from the
Department of Physics, University of California, Irvine Ca 92717, USA.}
and 
{\sl R. M. Williams}
$^{}$\footnote{Permanent address:
Department of Applied Mathematics and Theoretical Physics,
Wilberforce Road, Cambridge CB3 0WA, United Kingdom.} \\

\vspace{30pt}

Theory Division \\
CERN \\
CH-1211 Geneva 23, Switzerland \\

\vspace{20pt}

\end{center}

\begin{center} {\bf ABSTRACT } \end{center}

\noindent

Quantum corrections to the classical field equations, induced 
by a scale dependent gravitational constant, are analyzed in the case
of the static isotropic metric.
The requirement of general covariance for the resulting non-local
effective field equations puts severe restrictions on the nature of the
solutions that can be obtained. 
In general the existence of vacuum solutions to the effective field
equations restricts the value of the gravitational scaling
exponent $\nu^{-1}$ to be a positive integer greater than one.
We give further arguments suggesting that in fact only for $\nu^{-1}=3$
consistent solutions seem to exist in four dimensions.





\vfill

\pagestyle{empty}

\newpage

\pagestyle{plain}

\vskip 10pt

Over the last few years evidence has been increasing to suggest that quantum gravitation,
even though plagued by uncontrollable divergences in standard weak coupling
perturbation theory \cite{thooft},
might actually make sense, and lead to testable predictions at the non-perturbative level.
These new results in general arise from the non-trivial scaling properties of
the gravitational coupling constants in the vicinity of a non-trivial ultraviolet
fixed point in four dimensions.
As is often the case in physics, the best arguments do not come from
often incomplete and partial results in a single model,
but more appropriately from the level of consistency that various, often quite unrelated, 
field theoretic approaches provide.

The main aspect we wish to investigate in this paper is the nature of the
specific predictions about the running of Newton's constant $G$,
as they apply to the standard static isotropic metric.
Our starting point will be the solution of the non-relativistic
Poisson equation, which for a localized point source
can be investigated for various values of the gravitational scaling exponent $\nu$.
But a more appropriate setting will be a relativistic, generally covariant
framework, wherein the effects of the leading quantum correction can be studied
systematically, and for which we will show that the existence of vacuum solutions 
severely restricts the possible values for the exponent $\nu$.
Specifically, we will show that no consistent solution to the effective non-local
field equations can be found unless $\nu^{-1}$ is an integer greater than one.
To check the overall consistency of the results, a different
approach to the solution of the covariant effective field equations for
the static isotropic metric will be pursued, in terms of an effective
vacuum density and pressure.
In this case one finds that unless the exponent $\nu$ is equal to $1/3$,
a consistent solution cannot be obtained.

The starting point for our discussion is the form of the 
running gravitational coupling in the vicinity of the
ultraviolet fixed point at $G_c$, as obtained from the lattice theory of
gravity, and given in \cite{critical}
\beq
G(k^2) \; = \; G_c \left [ \; 1 \, 
+ \, a_0 \left ( { m^2 \over k^2 } \right )^{1 \over 2 \nu} \, 
+ \, O ( \, ( m^2 / k^2 )^{1 \over \nu} ) \; \right ]
\label{eq:grun_latt}
\eeq
with $m=1/\xi$, $a_0 > 0$ and $\nu \simeq 1/3$.
Usually the quantity $G_c$ in the above expression is identified with
the laboratory scale value, 
$ \sqrt{G_c} \sim \sqrt{G_{phys}} \sim 1.6 \times 10^{-33} cm$,
the reason being that the scale $\xi$ can be very large,
roughly of the same order as the scaled cosmological constant $\lambda$.
Quantum corrections on the r.h.s are therefore quite small as
long as $ k^2 \gg m^2 $, which in real space corresponds to the ``short distance''
regime $ r \ll \xi$.
\footnote{
The result of Eq.~(\ref{eq:grun_latt}) is 
in fact quite similar what one finds for gravity in $2+\epsilon$ 
dimensions \cite{eps,eps1,eps2}, if
one allows for a different value of exponent $\nu$ as one transitions from
two to four dimensions,
$ G(k^2) \; \simeq \; G_c \, \left ( \, 1 \, + \, 
\left ( m^2 / k^2 \right )^{(d-2)/2} \, + \, \dots \right ) $. 
See also the recent results discussed in \cite{reuter,litim}.}
For more details the reader is referred to the
recent papers [3-5], and further references therein.

For $k^2 \rightarrow 0$ the quantum correction proportional to $a_0$ diverges,
and the infrared divergence needs to be regulated.
A natural infrared regulator exists in the form of $m=1/\xi$, and
therefore a properly infrared regulated version of the previous expression is
\beq
G(k^2) \; \simeq \; G_c \left [ \; 1 \, 
+ \, a_0 \left ( { m^2 \over k^2 \, + \, m^2 } \right )^{1 \over 2 \nu} \, 
+ \, \dots \; \right ]
\label{eq:grun_m}
\eeq
with $m=1/\xi$ the (tiny) infrared cutoff.
Thus the gravitational coupling approaches the finite value 
$G_\infty = ( 1 + a_0 + \dots ) \, G_c $, independent of $m=1/ \xi$,
at very large distances $r \gg \xi$.
In this work we will be concerned with the static limit,
where the non-relativistic Newtonian potential can be defined as
\beq
\phi (r) \; = \; ( - M ) \int { d^3 {\bf k} \over (2 \pi)^3 } \, 
e^{i \, {\bf k} \cdot {\bf x} } \, G( {\bf k^2} ) \, { 4 \pi \over {\bf k^2} } 
\label{eq:pot_def}
\eeq
The static potential $\phi(r)$ can be obtained from Eq.~(\ref{eq:grun_m})
directly by Fourier transform, or equivalently from the solution
of Poisson's equation with a point source at the origin.
In the limit of weak fields, the relativistic field equations
give for the $\phi$ field (with $g_{00}(x) \simeq  -(1+2 \phi(x)\,)$
\beq
\Delta \, \phi(x) \; = \; 4 \, \pi \, G \, \rho (x) 
\label{eq:pois}
\eeq
and for a point source at the origin the first term on the r.h.s is just
$4 \pi M G \, \delta^{(3)} ( {\bf x}) $.
The solution for $\phi(r)$, obtained by Fourier transforming
back to real space Eq. (\ref{eq:pot_def}), gives in the large $r$ limit
\beq
\phi(r) \; \mathrel{\mathop\sim_{ r  \, \rightarrow \, \infty  }} \; 
- \, { M \, G \over r } \, \left [ 1 \, + \, a_0 \, 
\left ( 1 \, - \, c_l \, ( m \, r )^{ {1 \over 2 \nu} - 1 } \,
e^{- m r } \right ) \, \right ]
\label{eq:phi_large}
\eeq
with $c_l = 1 / ( \nu \, 2^{1 \over 2 \nu} \, \Gamma ( {1 \over 2 \nu } )) $. 
The part in $G(k^2)$ proportional to $a_0$ can equivalently be represented
as a source term $\rho_m $ in Poisson's equation, the latter
determined from the inverse Fourier transform of the correction term
in Eq.~(\ref{eq:grun_m}), 
\beq
a_0 \, M \, \left ( { m^2 \over {\bf k^2} \, + \, m^2 } \right )^{1 \over 2 \nu}
\eeq
One finds
\beq
\rho_m (r) \; = \; { 1 \over 8 \pi } \, c_{\nu} \, a_0 \, M \, m^3 \,
( m \, r )^{ - {1 \over 2} (3 - {1 \over \nu}) }  
\, K_{ {1 \over 2} ( 3 - {1 \over \nu} ) } ( m \, r ) 
\label{eq:rho_vac}
\eeq
with $ c_{\nu} \; \equiv \; 2^{ {1 \over 2} (5 - {1 \over \nu}) }
/ \sqrt{\pi} \, \Gamma( {1 \over 2 \, \nu} ) $.
Note that the vacuum polarization density $\rho_m (r) $ has the normalization property 
\beq
4 \, \pi \, \int_0^\infty \, r^2 \, d r \, \rho_m (r) \; = \;  a_0 \, M 
\label{eq:rho_vac2}
\eeq
and that $\rho_m(r)$ diverges at small $r$ for $\nu \geq 1/3$.
In the small $r$ limit and for general $\nu > {1 \over 3} $, one then 
finds from Poisson's equation, using the expansion of the modified Bessel
function $K_n (x)$ for small arguments,
\beq
\phi(r) \; 
\; \mathrel{\mathop\sim_{ r \, \rightarrow \, 0 }} \; 
- \, { M \, G \over r } \, + \,
a_0 \, M \, G \, c_s \, m^{1 \over \nu} \, r^{ { 1 \over \nu } - 1 } \, + \, \dots
\label{eq:phi_small}
\eeq
with $c_s = \nu \vert \sec \left ( { \pi \over 2 \nu } \right ) 
\vert / \Gamma ( {1 \over \nu } ) $. 

Solutions to Poisson's equation with a running $G$ provide some useful insights
into the structure of quantum corrections, but a complete analysis
requires a study of the full relativistic field equations, which
will be discussed next.
A set of effective field equations incorporating the running of
$G$ is obtained from the replacement \cite{effective} 
\beq
G \; \rightarrow \;  G( \Box) \; = \; G \, \left [ \, 1 \, + \, 
a_0 \, \left ( { m^2 \over \Box } \right )^{ 1 \over 2 \nu} \, 
+ \, \dots \; \right ] \; \equiv \; G \, \left ( \, 1 \, + \, A( \Box ) \, \right )
\label{eq:gbox}
\eeq
with the d'Alembertian $\Box$ expressing the running of $G$ as in 
either Eqs.~(\ref{eq:grun_latt}) or (\ref{eq:grun_m}).
The non-local effective field equations then read
\beq
R_{\mu\nu} \, - \, \half \, g_{\mu\nu} \, R \, + \, \lambda \, g_{\mu\nu}
\; = \; 8 \, \pi \, G  \, \left( 1 + A( \Box ) \right) \, T_{\mu\nu}
\label{eq:field}
\eeq
with $A( \Box ) $ given by Eq.~(\ref{eq:gbox}), and $\lambda \simeq 1 / \xi^2$.
The use of the d'Alembertian $\Box$ to describe the running of couplings in gauge theories
and quantum gravity was discussed in some detail, for example, in \cite{vilko}. 
The corresponding trace equation is
\beq
R \, - \, 4 \, \lambda \; = \; - \, 8 \, \pi \, G \, \left( 1 + A( \Box ) \right) \, T
\label{eq:field_tr}
\eeq
Being manifestly covariant, these expressions at least satisfy some
of the requirements for a set of consistent field equations
incorporating the running of $G$.
The d'Alembertian $\Box$ operator is defined here through 
the appropriate combination of covariant derivatives
\beq
\Box \; \equiv \; g^{\mu\nu} \, \nabla_\mu \nabla_\nu 
\eeq
and its explicit form depends on the specific tensor nature of
the object it is acting on.
In general the operator $A(\Box)$ has to be defined by a suitable analytic
continuation from positive
integer powers, which is usually done by computing $\Box^n$
for positive integer $n$, and then analytically continuing to 
$n \rightarrow -1/2\nu$.
Let us set for now the cosmological constant $\lambda=0$, since its
contribution can always be added at a later stage.
As long as one is interested in static isotropic solutions, one
can take for the metric the most general form
\beq
ds^2 \; = \; - \, B(r) \, dt^2 \, + \, A(r) \, d r^2 \, 
+ \, r^2 \, ( d \theta^2 \, + \, \sin^2 \theta \, d \varphi^2 )
\label{eq:metric}
\eeq
For the energy momentum tensor we will take the perfect fluid form 
\beq
T_{\mu \nu} \; = \; {\rm diag} \,
[ \, B(r) \, \rho(r), \, A(r) \, p(r) ,
\, r^2 \, p(r), \, r^2 \, \sin^2 \theta \, p(r) \, ]
\label{eq:perfect_si}
\eeq
and a point source as the origin is simply represented as
\beq
T_{\mu \nu} (r) \; = \; {\rm diag} [ \, B(r) \, \rho(r), \, 0, \, 0, \, 0 \, ]
\label{eq:enmom_point}
\eeq
with the source proportional to a $3-d$ delta function.

Consider first the trace equation
\beq
R \; = \; - 8 \, \pi \, G  (\Box) \, T 
\; \equiv \;  - \, 8 \, \pi \, G \, \left ( 1 \, + \, A( \Box ) \right ) \, T 
\; = \; + 8 \, \pi \, G \, \left ( 1 \, + \, A( \Box ) \right ) \, \rho 
\eeq
where we have used the fact that the point source at the origin
is described just by the density term.
One then computes the repeated action of the invariant d'Alembertian on $T$,
\beq
\Box \, (- 8 \pi G T ) \; = \; \Box (+ 8 \pi G \rho ) \; = \;
\frac{16\,G\,\pi \,\rho '}{r\,A} - 
  \frac{4\,G\,\pi \,A'\,\rho '}{{A}^2} + 
  \frac{4\,G\,\pi \,B'\,\rho '}{A\,B} + 
  \frac{8\,G\,\pi \,\rho ''}{A}
\eeq
In view of the rapidly escalating complexity of the problem, 
it seems sensible to expand around the Schwarzschild solution, and set
\beq
A(r)^{-1} \; = \; 1 \, - \, { 2 \, M \, G \over r} \, + \, { \sigma (r) \over r }
\;\;\;\;\;\;\;\;\;\;
B(r) \; = \; 1 \, - \, { 2 \, M \, G \over r} \, + \, { \theta (r) \over r }
\label{eq:sigma}
\eeq
where the correction to the standard solution are 
parametrized here by the two functions $\sigma(r)$ and $\theta(r)$, both
assumed to be ``small'', i.e. proportional to $a_0$ as in
Eq.~(\ref{eq:gbox}), with $a_0$ considered a small parameter.
To simplify the problem even further, we will assume that for 
$ 2 M G  \ll r \ll \xi $
(the ``physical'' regime) one can set
\beq
\sigma (r) \; = \; - \, a_0 \, M \, G \, c_\sigma \, r^\alpha 
\;\;\;\;\;\;\;\;\;\;
\theta (r) \; = \; - \, a_0 \, M \, G \,  c_\theta \, r^\beta 
\label{eq:sigma_p}
\eeq
This assumption is in part justified by the form of the non-relativistic
correction of Eqs.~(\ref{eq:phi_small}).
Then for $\alpha=\beta$ (the equations seem impossible to satisfy if
$\alpha$ and $\beta$ are different) one obtains for the scalar curvature
\beq
R \; = \; 0 \, + \, \alpha \, \left ( 2 \, c_\sigma + ( \alpha - 1 ) \, c_\theta \right ) \,
a_0 \, M \, G \, r^{\alpha-3} \, + \, + O( a_0^2 )
\label{eq:scalar_a0}
\eeq
A first result can be obtained in the following way.
Since in the ordinary Einstein case one has
for a perfect fluid $R = - 8 \pi G T = + 8 \pi G ( \rho - 3 p)$, and 
since $\rho_m (r) \sim r^{{1\over \nu} -3}$
from Eq.~(\ref{eq:rho_vac}) in the same regime, one concludes that a
solution is given by
\beq
\alpha \; = \; { 1 \over \nu } 
\label{eq:alpha_nu}
\eeq
which is also consistent with the Poisson equation result of Eq.~(\ref{eq:phi_small}).

The next step up would be the consideration of the action of $\Box$ on the point source,
as it appears in the full effective field equations of Eq.~(\ref{eq:field}), 
with again $T_{\mu\nu}$ described by Eq.~(\ref{eq:enmom_point}).
One perhaps surprising fact is the generation of an effective
pressure term by the action of $\Box$, suggesting that both terms should arise
in the correct description of vacuum polarization effects, 
\bea
\left( \Box \, T_{\mu\nu} \right )_{tt} \; & = & \; 
  - \frac{ \rho \,{B'}^2 }{2\,A\,B} + 
  \frac{2\,B\,\rho '}{r\,A} - 
  \frac{B\,A'\,\rho '}{2\,{A}^2} + 
  \frac{B'\,\rho '}{2\,A} + \frac{B\,\rho ''}{A}
\nonumber \\
\left( \Box \, T_{\mu\nu} \right )_{rr} \; & = & \; 
- \frac{ \rho \,{B'}^2 }{2\,{B}^2}
\label{eq:boxont}
\eea
and $ \left( \Box \, T_{\mu\nu} \right )_{\theta\theta} = 
\left( \Box \, T_{\mu\nu} \right )_{\varphi\varphi} = 0 $.
A similar effect, namely the generation of an effective vacuum pressure
term in the field equations by the action of $\Box$, was seen already
in the case of the Robertson-Walker \cite{effective}.

To check the overall consistency of the approach, consider next the set of
effective field equations that
are obtained when the operator $(1+A(\Box))$ appearing in 
Eqs.~(\ref{eq:field}) and (\ref{eq:field_tr}) is moved over to the
gravitational side.
Since the r.h.s of the field equations then vanishes for $r \neq 0$, one 
has apparently reduced the problem to one of finding vacuum solutions
of a modified, non-local field equation.
Let us first look at the relatively simple trace equation.
If we denote by $\delta R$ the lowest order variation (that is, of order $a_0$)
in the scalar curvature over the ordinary vacuum solution $R=0$, then one has
\beq
{ 1 \over 8 \pi G \, A( \Box ) } \; \delta R \; = \; 0
\label{eq:field_tr_rev1}
\eeq
On a generic scalar function $F(r)$ one has the following action of the
covariant d'Alembertian $\Box$ :
\beq
\Box \, F(r) \; = \; 
-\frac{A' F'}{2 A^2}+\frac{B' F'}
{2 A B}+\frac{2 F'}{r A}+\frac{F''}{A}
\eeq
Assuming a power law correction, as in Eq.~(\ref{eq:sigma_p}),
with $\alpha=\beta$, as in Eq.~(\ref{eq:scalar_a0}),
one then finds
\bea
\Box \,\; R \; & \rightarrow & \; a_0 \, M \, G \, r^{\alpha-5} \, 
\left( \, 2 \, c_\sigma + c_\theta \, ( \alpha - 1 ) \, \right ) \, 
\alpha \, (  \alpha - 2 ) \, ( \alpha -3 ) 
\nonumber \\
\Box^2 \, R \; & \rightarrow & \; a_0 \, M \, G \, r^{\alpha - 7} \, 
\left( \, 2 \, c_\sigma + c_\theta \, ( \alpha - 1 )  \, \right) \, 
\alpha \, ( \alpha  - 2 ) \, ( \alpha - 3 ) \, ( \alpha - 4 )
\, ( \alpha - 5 )
\eea
and so on, and for general $n \rightarrow + { 1 \over 2 \nu} $
\beq
\Box^n \, R \; \rightarrow \;
a_0 \, M \, G \, \left ( \, 2 \, c_\sigma + c_\theta \, ( \alpha - 1 ) \, \right) \, 
{ \alpha \, \Gamma ( 2 + {1 \over \nu } - \alpha ) \over \Gamma ( 2 - \alpha ) }
\; r^{\alpha - {1 \over \nu } - 3 } 
\eeq
Therefore the only possible power solution for $r \gg M G $
is $\alpha= 0, \, 2 \dots {1 \over \nu} + 1$, with $c_\sigma$ and $c_\theta$
unconstrained to this order.

Next we examine the full effective field equations (as opposed
to just their trace part) as in Eq.~(\ref{eq:field}) with $\lambda=0$,
If one denotes by 
$\delta G_{\mu\nu} \equiv \delta \left( R_{\mu\nu} - \half \, g_{\mu\nu} \, R \right)$
the lowest order variation (that is, of order $a_0$)
in the Einstein tensor over the ordinary vacuum solution $G_{\mu\nu}=0$, then one has
\beq
{ 1 \over 8 \pi G \, A( \Box ) } \;\,
\delta \left( R_{\mu\nu} - \half \, g_{\mu\nu} \, R \right) \; = \; 0
\label{eq:field_vac1}
\eeq
again for $r \neq 0$.
Here the covariant d'Alembertian operator $\Box$ acts on a second rank tensor 
and would thus seem to require the calculation of as many as 1920 terms,
of which many fortunately vanish by symmetry.
In the static isotropic case the components of the Einstein tensor are given by 
\bea
G_{tt} \; & = & \; \frac{A' B}{r A^2}-\frac{B}{r^2 A}+\frac{B}{r^2}
\nonumber \\
G_{rr} \; & = & \; -\frac{A}{r^2}+\frac{B'}{r B}+\frac{1}{r^2}
\nonumber \\
G_{\theta\theta} \; & = & \; -\frac{B'^2 r^2}{4 A B^2}-
\frac{A' B' r^2}{4 A^2 B}
+\frac{B'' r^2}{2 A B}
-\frac{A' r}{2 A^2}+\frac{B' r}{2 A B}
\nonumber \\
G_{\varphi\varphi} \; & = & \; \sin^2 \theta \, G_{\theta\theta} 
\eea
After acting with $\Box$ on this expression one finds a rather complicated result.
Here we will list only $ (\Box G)_{tt} $:  
\bea
&& \frac{6 B A'^3}{r A^5}+\frac{2 B A'^2}{r^2 A^4}
-\frac{4 B' A'^2}{r A^4}-\frac{2 B A'}{r^3 A^3}
-\frac{6 B A'' A'}{r A^4}+\frac{B'' A'}{r A^3}
+\frac{6 B}{r^4 A}
\nonumber \\
&& -\frac{6 B}{r^4 A^2}
-\frac{4 B'}{r^3 A}+\frac{4 B'}{r^3 A^2}
-\frac{B A''}{r^2 A^3}+\frac{2 B' A''}{r A^3}
+\frac{B''}{r^2 A}-\frac{B''}{r^2 A^2}
+\frac{B A^{(3)}}{r A^3}
\eea
If one again assumes that the corrections are given by a power, as in
Eq.~(\ref{eq:sigma_p}), with $\alpha=\beta$, then one has to lowest order
\bea
G_{tt} \; & = & \; 
a_0 \, M \, G \, c_\sigma \, \alpha \, r^{\alpha -3} 
\nonumber \\
G_{rr} \; & = & \; 
- \, a_0 \, M \, G \, (c_\sigma+c_\theta (\alpha -1)) \, r^{\alpha -3} 
\nonumber \\
G_{\theta\theta} \; & = & \; 
- \frac{1}{2} \, a_0 \, M \, G \, (c_\sigma+c_\theta (\alpha -1))
\, (\alpha -1) \, r^{\alpha -1} 
\eea
with the $\varphi\varphi$ component again proportional to the 
$\theta\theta$ component.
Applying $\Box$ on the above Einstein tensor one then gets
\bea
\left ( \Box \, G \right )_{tt} \; & = & \; 
a_0 \, G\,M\, c_\sigma \, \alpha  ( \alpha - 2) \, ( \alpha - 3) \, r^{\alpha - 5}
\nonumber \\
\left ( \Box \, G \right )_{rr} \; & = & \; 
-  \, a_0 \,G \,M \, \left( c_\sigma + c_\theta \, ( \alpha -1 )  \right) \,
\alpha \, ( \alpha - 3 ) \, r^{\alpha - 5 }
\nonumber \\
\left ( \Box \, G \right )_{\theta\theta} \; & = & \; 
- \, \frac{1}{2} \, a_0 \,G \,M \, \left( c_\sigma + c_\theta \, ( \alpha -1 )  \right) \,
\alpha \, ( \alpha - 3 )^2 \, r^{\alpha - 3 }
\eea
(with the $\varphi\varphi$ component proportional to the $\theta\theta$ component),
and so on. 
One then has for general $n \rightarrow + { 1 \over 2 \nu} $
\bea
\left ( \Box^n \, G \right )_{tt} \; & \rightarrow & \; a_0 \,G\,M\, c_\sigma \, 
\frac{\Gamma (2 + {1 \over \nu} - \alpha )}{\left( \alpha -1 \right) \, \Gamma (-\alpha )}
\, r^{\alpha - 3 - {1 \over \nu}}
\nonumber \\
\left ( \Box^n \, G \right )_{rr} \; & \rightarrow & \; -  a_0 \,G \,M \, 
\left( c_\sigma + c_\theta \, ( \alpha -1 )  \right) \,
\frac{ \Gamma (2 + {1 \over \nu} - \alpha )}{\left( \alpha -1 \right) \,
\left( \alpha - {1 \over \nu} \right) \, \Gamma(-\alpha )}
\, r^{\alpha - 3 - {1 \over \nu}  }
\nonumber \\
\left ( \Box^n \, G \right )_{\theta\theta} \; & \rightarrow & \; 
- \frac{1}{2} \, a_0 \,G \,M \, \left( c_\sigma + c_\theta \, ( \alpha -1 )  \right) \,
\frac{\left( \alpha - 1 - {1 \over \nu} \right) \, \Gamma (2 + {1 \over \nu} - \alpha )}
{\left( \alpha -1 \right) \, \left( \alpha - {1 \over \nu} \right) \, \Gamma(-\alpha )}
\, r^{\alpha - 1 - {1 \over \nu} }
\eea
Inspection of the above results reveals a common factor $1/\Gamma(-\alpha)$,
which would allow only integer powers $\alpha=0,1,2 \dots$, but the additional
factor of $1/(\alpha-1)$ excludes $\alpha=1$ from being a solution.
Even for $\alpha$ close to $1 / \nu$ (as expected on the basis of the non-relativistic
expression of Eq.~(\ref{eq:phi_small}), as well as from
Eq.~(\ref{eq:alpha_nu})) $\nu \sim 1/\alpha - \epsilon$ 
only integer values $\alpha=2,3,4 \dots $ are allowed.
In general the problem of finding a complete general solution to the effective
field equations by this method lies in the difficulty of computing
arbitrarily high powers of $\Box$ on general functions such as
$\sigma(r)$ and $\theta(r)$, which eventually involve a large number
of derivatives. 
Assuming for these functions a power law dependence on $r$ simplifies
the problem considerably, but also restricts the kind of solutions
that one is likely to find.
More specifically, if the solution involves (say for small $r$, but still 
with
$r \gg 2 M G$) a term of the type $r^\alpha \ln m r$, as in
Eqs.~(\ref{eq:phi_small}), (\ref{eq:a_small_r3}) and (\ref{eq:b_small_r3})
for $\nu \rightarrow 1/3$, then this method will have to be dealt
with very carefully. 
This is presumably the reason why in some of the $\Gamma$-function
coefficients encountered here one finds a power solution (in fact $\alpha=3$)
for $\nu$ close to a third, but one gets indeterminate expression
if one sets exactly $\alpha=1/\nu=3$.

The earlier discussion of the non-relativistic case suggests that the
quantum correction due to the running of $G$ can be approximately 
described by Poisson's equation, with a source term related to
a vacuum energy density $\rho_m(r)$, distributed around
the static source of strength $M$ in accordance with the result of 
Eqs.~(\ref{eq:rho_vac}) and Eq.~(\ref{eq:rho_vac2}).
These expressions, in turn, were obtained by Fourier transforming back 
to real space the original result for $G(k^2)$ of Eq.~(\ref{eq:grun_m}).
Furthermore, in the preceding discussion of the relativistic case
it was found (as in \cite{effective} for the Robertson-Walker metric case)
that a manifestly covariant implementation of the running of $G$, 
via the $G(\Box)$ given in Eq.~(\ref{eq:gbox}),
will induce a non-vanishing effective pressure term in the field equations.
This result can be seen clearly, in the case of the static isotropic metric,
for example from the result of Eq.~(\ref{eq:boxont}).
We will therefore now consider a relativistic perfect fluid, 
with energy-momentum tensor,
which in the static isotropic case reduces to Eq.~(\ref{eq:perfect_si}).
The $tt$, $rr$ and $\theta\theta$ components of the field equations then read 
\bea
-\lambda  B+\frac{A' B}{r A^2}
-\frac{B}{r^2 A}+\frac{B}{r^2} & = & 8 \pi G  \, B \, \rho 
\nonumber \\
\lambda  A-\frac{A}{r^2}+\frac{B'}{r B}+\frac{1}{r^2}
& = & 8 \pi G  \, A \, p 
\nonumber \\
\lambda  r^2
-\frac{B'^2 r^2}{4 A B^2}
-\frac{A' B' r^2}{4 A^2 B}
+\frac{B'' r^2}{2 A B}
-\frac{A' r}{2 A^2}
+\frac{B' r}{2 A B} & = & 8 \pi G \, r^2 \, p
\eea
with the $\varphi\varphi$ component equal to $\sin^2 \theta$ times the
$\theta\theta$ component.
Energy conservation $\nabla^{\mu} \, T_{\mu\nu} =0 $ implies
\beq
\left ( \, p+\rho  \, \right ) \frac{ B'}{2 B} + p' \; = \; 0
\eeq
and forces a definite relationship between $B(r)$, $\rho(r)$ and $p(r)$.
The three field equations and the energy conservation equation
are, as usual, not independent, because of the Bianchi identity.

It seems reasonable to attempt to solve the above equations (usually
considered in the context of relativistic stellar structure)
with the density given by the $\rho_m (r)$ of
Eqs.~(\ref{eq:rho_vac}).
This of course raises the question of how the relativistic pressure $p(r)$
should be chosen, an issue that the non-relativistic calculation
did not have to address.
We will argue below that covariant energy conservation completely
determines the pressure in the static case, leading to
consistent equations and solutions (note that in particular it would
not be consistent to take $p(r)=0$). 
Since the function $B(r)$ drops out of the $tt$ field equation,
the latter can be integrated immediately, giving
\beq
A(r)^{-1} \; = \; 1 \, + \, { c_1 \over r } \, - \, { \lambda \over 3 } \, r^2 
\, - \, { 8 \pi G \over r } \, \int_0^r d x \, x^2 \, \rho (x)
\label{eq:ar}
\eeq
It also seems natural here to identify $c_1 = - 2 M G $,
which of course corresponds to the correct solution for $a_0=0$ ($p=\rho=0$).
Next, the $rr$ field equation can be solved for $B(r)$,
\beq
B(r) \; = \; \exp \left \{ c_2 \, - \, \int_{r_0}^r \, d y \, 
\frac{1 + A(y) \left( \lambda \, y^2 - 8 \pi G \, y^2 \, p(y) - 1 \right)}{y} 
\right \}
\label{eq:br}
\eeq
with the constant $c_2$ again determined by the requirement that the
above expression for $B(r)$ reduce to the standard Schwarzschild solution
for $a_0=0$ ($p=\rho=0$), giving $c_2=\ln (1 - 2 M G / r_0 - \lambda r_0^2 / 3)$.
The last task left therefore is the determination of the pressure $p(r)$.
Using the $rr$ field equation, $B'(r)/B(r)$ can be expressed in
term of $A(r)$ in the energy conservation equation.
Inserting then the explicit expression for $A(r)$, from Eq.~(\ref{eq:ar}), one obtains
\beq
p'(r) + \frac{\left ( 8 \pi G \, r^3 \, p(r) \, + \, 2 M G \, 
- \, { 2 \over 3} \lambda r^3 \, 
+ \, 8 \pi G \, \int_{r_0}^r dx \, x^2 \rho (x) \right ) \, (p(r)+\rho (r))}
{2 \, r \left( r \, - \, 2 M G \, - \, {\lambda \over 3}\, r^3  
- 8 \pi G \, \int_0^r dx \, x^2 \, \rho (x) \right) } \; = \; 0
\label{eq:pres1}
\eeq
which is usually referred to as the equation of hydrostatic equilibrium.
From now on we will focus only the case $\lambda=0$.
The last equation, a non-linear differential equation for $p(r)$, can
be solved to give the desired solution $p(r)$, which then,
by equation Eq.~(\ref{eq:br}), determines the remaining function $B(r)$.
In our case though it will be sufficient to solve the above equation
for small $a_0$, where $a_0$
(see Eqs.~(\ref{eq:grun_m}) and (\ref{eq:rho_vac}))
is the dimensionless parameter which,
when set to zero, makes the solution revert back to the classical one.
It will also be convenient to pull out of $A(r)$ and $B(r)$ the
Schwarzschild solution part, by introducing the small corrections
$\sigma(r)$ and $\theta(r)$, as defined in Eq.~(\ref{eq:sigma}),
both of which are expected to be proportional to the parameter $a_0$.
One then has 
\beq
\theta (r) \; = \; 
\exp \left \{ c_2 + \int_{r_0}^r dy \, { 1 + 8 \pi G \, y^2 \, p(y) \over
y - 2 M G - 8 \pi G \int_0^y dx \, x^2 \, \rho (x) } \right \}
\, + \, 2 M G \, - \, r 
\label{eq:theta_sol}
\eeq
Again, the integration constant $c_2$ needs to be chosen here so that the normal
Schwarzschild solution is recovered for $p=\rho=0$.  
To order $a_0$ the resulting equation for $p(r)$, from Eq.~(\ref{eq:pres1}), is
\beq
\frac{M G \, ( p(r) \, + \, \rho (r))} {r \, ( r \, - \, 2 M G )} 
\, + \, p'(r) \; \simeq \; 0
\label{eq:pres0}
\eeq
Note that in regions where $p(r)$ is slowly varying, $p'(r) \simeq 0$,
one has $p \simeq -\rho $, i.e. the fluid contribution is
acting like a cosmological constant term with 
$ \sigma (r) \sim \theta (r) \sim - (\rho /3) \, r^3$.
The last differential equation can then be solved for $p(r)$,
\beq
p_m (r) \; = \; {1 \over \sqrt{1 - \frac{2 M G}{r} } } \,
\left ( c_3 - \int_{r_0}^r dz \, 
\frac{M G \rho (z)} {z^2 \, \sqrt{1-\frac{2 M G}{z}} }
\right )
\label{eq:pres0_sol}
\eeq
where the constant of integration has to be chosen so that
when $\rho(r)=0$ (no quantum correction) one has $p(r)=0$ as well.
Because of the singularity in the integrand at $r= 2 M G$, we will
take the lower limit in the integral to be $r_0 = 2 M G + \epsilon$,
with $\epsilon \rightarrow 0$.
To proceed further, one needs the explicit form for $\rho_m(r)$, which
was given in Eqs.~(\ref{eq:rho_vac}) and (\ref{eq:rho_vac2}).
The required integrands involve for general $\nu$ the modified Bessel function $K_n(x)$,
and can be therefore a bit complicated.
Here we will limit our investigation to the small $r$ ($ m r \ll 1$)
and large $r$ ($m r \gg 1$) behavior.
Since $m=1/\xi$ is very small, the first limit appears to be of greater physical
interest.
For small $r$ the density $\rho_m(r)$ has the following behavior
(see Eq.~(\ref{eq:rho_vac})),
\beq
\rho_m (r) \; \mathrel{\mathop\sim_{ r \, \rightarrow \, 0 }} \; 
A_0 \; r^{ {1 \over \nu} - 3 } 
\label{eq:rho_small}
\eeq
for $\nu > 1/3$, with the constant
\beq
A_0 \; = \; { \vert \sec \left ( { \pi \over 2 \nu } \right ) \vert 
\over 4 \pi \, \Gamma \left ( {1 \over \nu} -1 \right ) } \, a_0 \, M \, m^{1 \over \nu}
\eeq
determined from the small $x$ behavior of the modified Bessel function
$K_n(x)$.
For $\nu < 1/3$ $\rho_m (r) \, \sim \, {\rm const.} \, a_0 \, M \, m^3 $, 
independent of $r$.
For $\nu=1/3$ the expression for $\rho_m(r)$ is given later in Eq.~(\ref{eq:rho_vac_3}).
Therefore in this limit, with ${1 \over 3} < \nu < 1 $, one has 
\beq
A^{-1}(r) \; = \; 1 \, - \, { 2 \, M \, G \over r} \, - \, 
2 \, a_0 \, M \, G \, c_s \, m^{1 \over \nu} \, r^{{1\over\nu}-1} \, + \, \dots
\label{eq:a_small_r}
\eeq
with the constant $c_s = \nu \vert \sec \left ( { \pi \over 2 \nu } \right )
\vert / \Gamma ( {1 \over \nu } - 1 ) $.
For $\nu=1/3$ the last contribution is indistinguishable from
a cosmological constant term $- {\lambda \over 3} r^2$, except 
for the fact that the coefficient here is quite different, 
being proportional to $\sim a_0 \, M \, G \, m^3$.
To determine the pressure, we suppose that it as well has a power
dependence on $r$ in the regime under consideration, 
$p_m (r) = c_p \, A_0 \, r^\gamma $, where $c_p$ is a numerical constant,
and then substitute $p_m (r)$ into the pressure equation Eq.~(\ref{eq:pres0}).
This gives, past the horizon $r \gg 2 M G$,
\beq
( 2 \gamma \, - \,1 ) \, c_p \, M \, G \, r^{\gamma - 1} \, 
- \, c_p \, \gamma \, r^\gamma \, - \, M \, G \, r^{1/\nu -4} \, \simeq 0
\eeq
giving the same power $\gamma=1/\nu-3$ as for $\rho(r)$, $c_p=-1$
and surprisingly also $\gamma=0$, implying that in this regime
only $\nu=1/3$ gives a consistent solution.
Again, the resulting correction is quite similar to what one would expect
from a cosmological term, with an effective 
$\lambda_m / 3 \simeq  8 \pi \, \nu \, a_0 \, M \, G \, m^{1 \over \nu} $.

The case $\nu=1/3$ requires special treatment, and
one needs to go back to the expression for $\rho_m(r)$ for $\nu=1/3$,
\beq
\rho_m (r) \; = \; {1 \over 2 \pi^2 } \, a_0 \, M \, m^3 \, 
\, K_0 ( m \, r ) 
\label{eq:rho_vac_3}
\eeq
For small $r$ one then has
\beq
\rho_m (r) \; = \; - \, {a_0 \over 2 \pi^2 } \, M \, m^3 \, 
\left ( \ln { m \, r \over 2 } \, + \, \gamma \, \right ) \, + \, \dots
\label{eq:rho_vac_3_s}
\eeq
and consequently from Eq.~(\ref{eq:ar}),
\beq
A^{-1} (r) \; = \; 1 \, - { 2 \, M \, G \over r } \, + \, 
{4 \, a_0 \, M \, G \, m^3 \over 3 \, \pi } \, r^2 \, \ln \, ( m \, r ) 
\, + \, \dots
\label{eq:a_small_r3}
\eeq
From Eq.~(\ref{eq:pres0}) one then obtains an expression for the
pressure $p_m(r)$, and one finds
\beq
p_m (r) \; = \; 
\frac{a_0 M m^3 \log (m r) }{2 \pi ^2}
-\frac{a_0 M m^3 \log \left (r + r \sqrt{1 - { 2 M G \over r } } - M G \right) }
{2 \pi ^2 \sqrt{1- { 2 M G \over r } }}
+\frac{a_0 M m^3}{\pi ^2}
+\frac{a_0 M m^3 c_3 }{2 \pi^2 \sqrt{1- {2 M G \over r } }}
\eeq
where $c_3$ is again an integration constant.
Here we will be content with the $r \gg 2 M G $ limit of the above expression,
which we shall write therefore as 
\beq
p_m (r) \; = \; {a_0 \over 2 \pi^2 } \, M \, m^3 \, \ln \, (  m \, r ) 
\, + \, \dots
\label{eq:p_vac_3}
\eeq
After performing the required $r$ integral in Eq.~(\ref{eq:theta_sol}),
and evaluating the resulting expression in the limit $r \gg 2 M G$, one obtains
an expression for $\theta (r)$, and from it
\beq
B(r) \; = \; 1 \, - \, { 2 \, M \, G \over r} \, + \, 
{4 \, a_0 \, M \, G \, m^3 \over 3 \, \pi } \, r^2 \, \ln \, ( m \, r ) 
\, + \, \dots
\label{eq:b_small_r3}
\eeq
The expressions for $A(r)$ and $B(r)$ are, for $r \gg 2 \, M \, G$, consistent with
a gradual slow increase in $G$ in accordance with the formula
\beq
G \; \rightarrow \; G(r) \; = \; 
G \, \left ( 1 \, + \, 
{ a_0 \over 3 \, \pi } \, m^3 \, r^3 \, \ln \, { 1 \over  m^2 \, r^2 }  
\, + \, \dots
\right )
\label{eq:g_small_r3}
\eeq
and therefore consistent as well with the original result of Eqs.~(\ref{eq:grun_latt})
or (\ref{eq:grun_m}), namely that the classical laboratory value of $G$ 
is obtained for $ r \ll \xi $.
In fact it is quite reassuring that the renormalization properties of $G(r)$
as inferred from $A(r)$ are the same as what one finds from $B(r)$.
For large $r$ one has instead, from Eq.~(\ref{eq:rho_vac}) for $\rho_m(r)$,
\beq
\rho_m (r) \; \mathrel{\mathop\sim_{ r \, \rightarrow \, \infty }} \; 
A_0 \; r^{ {1 \over 2 \, \nu} - 2 } \, \, e^{- \, m \, r} 
\label{eq:rho_large}
\eeq
with $A_0 = 1 / \sqrt{128 \pi} \, c_{\nu} \, a_0 \, M \, m^{1 + { 1 \over 2 \nu }} $.
In the same limit, the integration constants is chosen so that the
solution for $A(r)$ and $B(r)$ at large $r$ corresponds to a mass $M' = (1+a_0) M$
(see the expression for the integrated density in Eq.~(\ref{eq:rho_vac2})),
or equivalently
\beq
\sigma (r) \; \sim \; \theta (r) 
\; \mathrel{\mathop\sim_{ r \, \rightarrow \, \infty }} \; - 2 \, a_0 \, M \, G
\eeq
One then recovers a result similar to the non-relativistic expression of
Eq.~(\ref{eq:phi_large}), with $G(r)$ approaching the constant value
$G_{\infty} = (1 + a_0) \, G$, up to an exponentially small correction 
in $ m r $ at large $r$.

In conclusion, it appears that a solution to the relativistic static isotropic
problem of the running gravitational constant can be found, provided that
the exponent $\nu$ in either Eq.~(\ref{eq:grun_m}) or Eq.~(\ref{eq:field})
is close to one third.
This last result seems to be linked with the fact that the running coupling
term acts in some way like a local cosmological constant term, for which the
$r$ dependence of the vacuum solution for small $r$ is fixed by the nature
of the Schwarzschild solution with a cosmological constant term.
Furthermore, in $d \ge 4$ dimensions the Schwarzschild solution to Einstein
gravity with a cosmological term is given by \cite{perry}
\beq
A^{-1}(r) \; = \; B(r) \; = \; 1 - 
{ 8 \, M \, G \, \pi \, \Gamma({d-1 \over 2}) \over (d-2) \, \pi^{d-1 \over 2} }
\, r^{3-d} - { 2 \lambda \over (d-2)(d-1) } \, r^2
\eeq
which would suggest, in analogy with the results for $d=4$ given previously,
that in $d \ge 4 $ dimensions only $\nu=1/(d-1)$ is possible, if the
correction again behaves locally like a cosmological constant term.
This last result would also be in agreement with the exact value
$\nu=0$ found at $d=\infty$ \cite{larged}. 

To summarize, the starting point for our discussion of the renormalization group
running of $G$ is Eq.~(\ref{eq:grun_latt}), valid at short distances $k \gg m$,
or its improved infrared regulated version of Eq.~(\ref{eq:grun_m}).
While a solution to the non-relativistic Poisson equation
can be given for various values of the exponent $\nu$,
the scale dependence of $G$ can also be consistently embedded
in a relativistic covariant framework using the d'Alembertian $\Box$ operator.
This then leads to a set of nonlocal effective field equations, whose
consequences can be worked out for the static isotropic
metric, at least in a regime where $ 2 M G \ll r \ll \xi$, and under the assumption
of a power law correction.
We have found that the structure of the leading quantum correction is such that it
severely restricts the possible values for the exponent $\nu$, in the sense that 
no consistent solution to the effective non-local
field equations, incorporating the running of $G$, can be found unless $\nu^{-1}$
is an integer.
A somewhat different approach to the solution of the static isotropic metric
was pursued in terms of an effective vacuum density
of Eq.~(\ref{eq:rho_vac}), and a vacuum pressure chosen so as to satisfy a
covariant energy conservation for the vacuum polarization contribution.
The main result there is the derivation, from the relativistic field equations, of
an expression for the metric coefficients $A(r)$ and $B(r)$, given for 
$ 2 M G \ll r \ll \xi$ in Eqs.~(\ref{eq:a_small_r3}), (\ref{eq:b_small_r3})
and (\ref{eq:g_small_r3}).
From the nature of the solution for $A(r)$ and $B(r)$ one finds again 
that unless the exponent $\nu$ is close to $1/3$, a consistent solution
to the field equations cannot be found.

\vspace{20pt}

{\bf Acknowledgements}

The authors wish to thank Luis \'Alvarez Gaum\'e, Gabriele Veneziano and the
Theory Division at CERN for their warm hospitality.
The work of Ruth M. Williams was supported in part by the UK Particle
Physics and Astronomy Research Council.


\vfill

\newpage

\end{document}